\documentclass[final,10p,times,twocolumn]{elsarticle}
\usepackage[T1]{fontenc}
\usepackage[utf8]{inputenc}
\usepackage{amsmath,amssymb,amsthm}
\usepackage{bm}
\usepackage{algorithm}
\usepackage{algpseudocode}
\usepackage{graphicx}
\usepackage{booktabs}
\usepackage{multirow}
\usepackage[numbers]{natbib}
\usepackage{url}
\usepackage[hidelinks]{hyperref}
\usepackage{xcolor}
\usepackage{tikz}
\usetikzlibrary{arrows.meta,positioning}

\newtheorem{theorem}{Theorem}
\newtheorem{proposition}{Proposition}

\newtheorem{corollary}{Corollary}
\newtheorem{assumption}{Assumption}
\newtheorem{definition}{Definition}

\DeclareMathOperator*{\argmax}{arg\,max}

\newcommand{\Ztx}{Z_{\mathrm{tx}}}
\newcommand{\Zrx}{Z_{\mathrm{rx}}}
\newcommand{\Itx}{I_{\mathrm{tx}}}
\newcommand{\Irx}{I_{\mathrm{rx}}}
\newcommand{\Lenc}{L_{\mathrm{enc}}}
\newcommand{\Lch}{L_{\mathrm{ch}}}
\newcommand{\Pe}{P_{\mathrm{e}}}
\newcommand{\Aenc}{\mathcal{A}_{\mathrm{enc}}}
\newcommand{\Ach}{\mathcal{A}_{\mathrm{ch}}}
\newcommand{\Ihatup}{\hat{I}^{\uparrow}_{\mathrm{rx}}}
\newcommand{\Ihatdn}{\hat{I}^{\downarrow}_{\mathrm{rx}}}
\newcommand{\Itxhatup}{\hat{I}^{\uparrow}_{\mathrm{tx}}}
\newcommand{\Itxhatdn}{\hat{I}^{\downarrow}_{\mathrm{tx}}}
\newcommand{\Yset}{\mathcal{Y}}

\newcommand{\figorbox}[3]{%
  \IfFileExists{#1}{\includegraphics[width=#2]{#1}}%
  {\setlength{\fboxsep}{6pt}\fbox{\parbox[c][2.6cm][c]{\dimexpr#2-14pt\relax}{\centering
   \footnotesize\itshape #3\\[2pt]\textup{(upload \texttt{\detokenize{#1}})}}}}}

\begin{document}

\begin{frontmatter}

\title{When Cause Attribution Is Not Enough: Action-Value Control for Task-Oriented Semantic Communication with Discrete Encoder Switching}

\author[1]{Moirangthem Tiken Singh\corref{cor1}}
\ead{tiken.m@dibru.ac.in}
\affiliation[1]{organization={Department of Computer Science and Engineering,
Dibrugarh University Institute of Engineering and Technology (DUIET),
Dibrugarh University}, addressline={Dibrugarh, Assam}, country={India}}
\cortext[cor1]{Corresponding author}

\begin{abstract}
A task-oriented semantic communication receiver that selects among independently trained encoders must answer two questions when performance degrades: which component is responsible, and which available action best restores task performance. These are distinct decision problems. Under discrete encoder switching, correctly identifying the dominant information-loss component does not necessarily identify the action that most reduces task error. Diagnosis and intervention should therefore be treated as separate problems. We support this finding with two complementary instruments. The first is a receiver-side diagnostic that partitions task-information loss into encoder and channel components and supplies population-valid task-error bounds conditional on one-sided information estimators. The second is an offline-supervised, online label-free action-value controller built from two monotone (isotonic) maps that target a reduction in measured error. The main experiments focus on two complementary axis-aligned actions that cleanly isolate the effect of discrete switching; a pilot further examines a combined action. Evaluation covers a controlled model with closed-form ground truth, three DeepJSCC vision systems, and a non-vision pilot. On STL-10, the diagnosis correctly identifies the larger loss component at 80\% of operating points, yet matches the oracle-best action only 25\% of the time. Structural mismatch caps any diagnosis-driven loss-ordering policy at $46\%$ accuracy, even with perfect knowledge of the ordering. The action-value controller, which exploits the magnitude of each loss deficit, achieves 0.86, compared to 0.28 for the raw rule. Where the encoder is rarely the bottleneck, the controller naturally retains the physical-layer default, yielding robust and selective intervention.
\end{abstract}

\end{frontmatter}

\section{Introduction}
\label{sec:intro}
Semantic and task-oriented communication transmits only the information a downstream task requires. This approach can substantially reduce overhead in machine-centric Internet of Things (IoT) and sixth-generation (6G) systems \cite{gunduz2023,xie2021,zhang2021}. Deep joint source--channel coding (DeepJSCC) realises the idea by training the transmitter and receiver jointly \cite{bourtsoulatze2019}. Most monitoring tools for these systems, however, still track bits, reconstruction quality, or channel quality. None of these quantities directly answers the two questions that matter most for deployment: how much task-relevant information actually arrived, and, when performance drops, which fix helps the most.

We identify three distinct gaps. The first is measurement. Existing task-information proxies \cite{stii,cosmic} typically do not bound the downstream task error, and they are generally not computable online without labels. The second is attribution. When performance drops, the receiver should determine whether the encoder or the channel is responsible. Importance-based and adaptive-rate methods \cite{ciba,entropyrate2025,gpcr} guide transmission decisions but do not directly identify which part of the system is at fault. Signal-to-noise ratio (SNR) alone cannot separate the two causes either. The third gap is the choice of intervention. Real systems switch among a small, fixed set of trained encoders rather than tuning a continuous parameter. Consequently, knowledge of which part of the system is more impaired does not automatically identify the action that reduces task error the most.

This third gap is the central concern of the present work. It creates a separation between diagnosis (``what is causing most of the loss?'') and intervention (``which available action helps the most?''). When control is continuous, the two questions tend to share the same answer. When encoders are switched in discrete jumps, they need not. One action can produce a larger drop in error than another even when its associated loss term is the smaller of the two. Perfect knowledge of the loss components therefore does not, in general, guarantee selection of the truly best discrete action. The central claim is precisely this: under discrete encoder switching, correctly identifying the dominant information-loss component does not necessarily identify the action that most reduces task error. Everything that follows is designed either to establish that claim or to act upon it.

The primary experiments investigate two complementary axis-aligned interventions: either increasing power while keeping the encoder fixed, or switching to a higher-rate encoder at fixed power, thereby allowing the discrete-switching behavior to be clearly isolated. An additional pilot study explores a combined intervention and shows that the separation between diagnosis and action remains informative even when both factors change simultaneously.

We therefore develop a receiver-side framework that keeps the two problems separate. The diagnostic stage divides task information loss into encoder loss and channel loss. It uses the deployed decoder to estimate predictive information and provides population-valid task-error bounds, provided that the estimators correctly bound the true information quantities. The control stage then learns how each possible action reduces task error. This mapping is trained offline using a labelled calibration set and does not require labels during deployment. The controller therefore selects actions based on their expected error reduction rather than simply choosing the action associated with the larger estimated loss.

Empirically, the true loss ordering can disagree sharply with the ordering of actions by their actual error reduction. Correctly identifying the larger loss component does not necessarily identify the action that produces the greatest reduction in task error. We explain this gap by partitioning action-selection error into a structural component, which remains even under perfect loss estimation, and an estimation component, which can be reduced through improved measurement. The structural component imposes a ceiling on diagnosis-driven loss-ordering policies specifically; it does not limit controllers that exploit the magnitude of the loss deficit. We therefore target error reduction directly using a simple isotonic (monotone) action-value learner.

\textbf{Contributions.}
\begin{enumerate}
\item \textbf{Separation of diagnosis from intervention, with quantification of the gap.} We partition action-selection error into a structural component and an estimation component. The partition shows precisely when improved attribution alone cannot improve the control decision. This separation is the primary contribution; the remaining items supply the mechanisms by which the separation is established and acted upon.
\item \textbf{Label-free, receiver-side diagnosis with population-level error bounds.} A decoder-derived diagnostic partitions task-information loss into encoder and channel parts. The associated bounds are population-valid conditional on the one-sided estimators being valid bounds on the true population quantities. The construction combines classic information-theoretic bounds with a data-processing decomposition.
\item \textbf{Offline-supervised, online label-free action-value controller.} Two isotonic maps estimate the realised error reduction of each action rather than relying on a simple loss-magnitude comparison. The maps retain magnitude information that a plain cause-attribution rule discards, while the certified diagnostic layer continues to supply its own population-valid commitments.
\item \textbf{Evaluation on a controlled model and learned vision systems.} The controlled model provides closed-form ground truth for validating the proposed diagnostic framework. The learned systems provide empirical evaluation across different operating conditions and calibration budgets.
\end{enumerate}

The framework combines several classic tools: Fano-type bounds~\cite{cover2006}, the Hellman--Raviv bound~\cite{hellman1970}, the Barber--Agakov variational bound~\cite{barber2003}, temperature scaling, a contrastive log-ratio (CLUB) bound~\cite{cheng2020}, isotonic regression, and data-processing arguments. Together these form a label-free, receiver-side framework for diagnosis and control. The classic distinction between source distortion and channel distortion in joint source--channel coding \cite{cover2006} is extended here to task-relevant information for the purpose of receiver-side diagnosis. Existing importance-attribution methods \cite{ciba} and conformal reliability methods \cite{cohen2023,angelopoulos2023} typically do not jointly address the separation of encoder-caused from channel-caused degradation and the certification of its source. Learned action-value rate selection \cite{gpcr} chooses interventions but offers neither the task-error certification nor the component-level attribution developed here.

\section{Related Work}
\label{sec:related}
Semantic communication research has shifted wireless system design away from bit-accurate delivery and toward the receiver's actual purpose \cite{strinati2021,gunduz2023}. In machine-centric traffic the value of a message lies entirely in what it enables a downstream model to accomplish. DeepJSCC made the idea practical: it trains transmitter and receiver jointly and degrades gracefully under conditions in which separated coding schemes fail completely \cite{bourtsoulatze2019}. Subsequent work extended the idea from reconstruction to inference, employing an information bottleneck to retain only what the task requires \cite{tishby1999,shao2022,xie2021}. This body of work established two facts: a representation can be compressed substantially without harming the task, and the encoder and the channel together determine what survives. It did not, however, address how a deployed receiver should monitor that survival. Although the training objective is task-aware, runtime monitoring continued to rely on reconstruction fidelity or physical-layer quality; neither indicates whether the task can still be performed.

A second line of work introduced measures that capture task relevance more directly. Task-information proxies and semantic similarity scores were proposed as alternatives to bit error rate (BER) for evaluating semantic links \cite{stii,cosmic}. These measures assess the meaning delivered by the link rather than the accuracy of individual symbols. However, two gaps remain. First, they do not directly quantify the task error that matters to the operator. Their relationship with task error is often assumed or observed to be monotonic, but it is rarely bounded from both sides. Therefore, a given proxy value does not provide a direct performance guarantee. Second, these measures are often evaluated using ground truth available in the laboratory. A deployed receiver, however, has neither the original source signal nor task labels. Such estimators are therefore difficult to use in practice. We address the first gap by estimating delivered task-relevant information at the receiver. The estimate is population-conditional and is computed from the deployed decoder without requiring online labels.

A separate line of work renders semantic systems adaptive by varying transmitted rate or allocated power according to channel conditions. Some schemes allow the transmitted description to expand or contract \cite{kurka2019,kurka2021,kurka2020}; others tie rate to channel quality and feature statistics \cite{bourtsoulatze2019,entropyrate2025}. Importance-attribution methods rank the parts of a representation that matter most and protect them accordingly \cite{ciba}. More recent work learns a surrogate model of task performance as a function of channel quality and compression setting, then selects a setting that meets a performance target \cite{gpcr}. These schemes typically share a common limitation: their control signal is a function of channel state, feature statistics, or decoder confidence, and each of these three quantities can rise under either failure mode (encoder-caused or channel-caused). A signal that cannot distinguish the two causes cannot choose between the two different remedies they require.

A fourth line of work applies distribution-free risk control to wireless decision-making. Scheduling, model selection, and resource allocation can be equipped with finite-sample reliability guarantees in place of average-case assurances \cite{cohen2023,angelopoulos2023}. This is valuable because it replaces informal calibration arguments with guarantees that hold without distributional assumptions, and any deployable system should be compatible with it. Its scope is narrow by design, however: a coverage guarantee certifies only that an error target is met, or reports when it is not. It does not by itself indicate which of several possible resources would correct the problem, because a guarantee of this kind is cause-agnostic by construction.

Taken together, these four lines of work leave a specific gap. Task-oriented design shows that the encoder and channel jointly determine receiver performance. Metric-based approaches measure this performance but generally do not certify it or estimate it without labels. Adaptive-control methods act on signals that do not directly identify the responsible factor. Reliability methods provide guarantees but do not identify the cause of degradation.

We address this gap in two steps. First, we develop a population-conditional diagnostic that uses the deployed decoder and requires no online labels. Second, we examine whether the diagnostic correctly identifies the best available action. This distinction is important because the receiver switches between a small set of trained encoders. The action with the larger estimated loss is not necessarily the action that produces the largest error reduction. When the diagnosis and the best action disagree, we use an action-value learner. It is trained offline on a labelled calibration split and operates without labels online. The learner predicts the error reduction produced by each action rather than relying on the ordering of loss magnitudes.

Our experiments quantify how often the diagnosis disagrees with the best action and evaluate the learner specifically in these cases. We focus on axis-aligned actions to isolate and clearly evaluate the discrete-switching effect.

\section{Methodology}
\label{sec:method}
A semantic link can fail in two structurally different ways. Either the encoder discards task-relevant information during compression, or the channel destroys information that the encoder had preserved. These two failure modes call for opposite remedies. Channel-caused loss can be mitigated by additional physical resources: power, bandwidth, or channel uses. Encoder-caused loss cannot be mitigated in this way; it requires a different encoder. Spending power on an encoder-limited link wastes budget, and retraining the encoder does not repair a channel-limited link. A controller that selects interventions by predicted error reduction, under a budget of one action per decision, therefore requires three elements: a label-free, receiver-side measure of delivered task information together with a task-error guarantee; an attribution of any shortfall to the encoder or the channel; and a rule that converts this attribution into a resource choice that remains valid when the encoder can be changed only over a small, discrete set of options.

\begin{figure}[h!]
\centering
\begin{tikzpicture}[
  font=\scriptsize, >={Latex[length=1.5mm]},
  proc/.style={draw, rounded corners=2pt, align=center, inner sep=3pt},
  tag/.style={font=\tiny\itshape, text=black!55, align=center},
  arr/.style={->, semithick},
]
\node[proc, text width=.82\columnwidth] (info)
  {\textbf{Delivered task information}\\
   measured at the receiver, from the deployed decoder, without labels};

\node[proc, text width=.82\columnwidth, below=3mm of info] (dec)
  {\textbf{Split the shortfall} into an encoder part and a channel part};
\draw[arr] (info) -- (dec);

\node[proc, text width=.37\columnwidth, below=10mm of dec, xshift=-.22\columnwidth]
  (diag) {\textbf{Diagnosis}\\ which part dominates?\\ (commit or abstain)};
\node[proc, text width=.37\columnwidth, below=10mm of dec, xshift=.22\columnwidth]
  (val)  {\textbf{Action selection}\\ which fix reduces\\ task error most?};
\draw[arr] (dec.south) -- (diag.north);
\draw[arr] (dec.south) -- (val.north);

\node[tag, below=0.5mm of diag, text width=.37\columnwidth]
  {guaranteed when it\\ commits};
\node[tag, below=0.5mm of val, text width=.37\columnwidth]
  {calibrated offline,\\ label-free online};

\draw[<->, dashed] (diag.east) -- node[font=\tiny, align=center, fill=white]
  {need not\\ agree} (val.west);

\node[proc, text width=.37\columnwidth, below=12mm of val] (fuse)
  {\textbf{Intervene selectively}\\ (default to physical layer\\ unless a fix clearly helps)};
\node[proc, text width=.37\columnwidth, below=3mm of fuse] (act)
  {\textbf{Chosen action}\\ switch encoder \emph{or}\\ add power};
\draw[arr] (val) -- (fuse);
\draw[arr] (fuse) -- (act);
\end{tikzpicture}
\caption{Overview of the framework. A single receiver-side measurement of
delivered task information is partitioned into an encoder part and a channel part.
These two parts feed two separate decisions: a diagnosis of which part
is responsible, and a choice of action that most reduces task error.
Under the stated one-sided-bound conditions, the diagnosis is guaranteed
correct wherever it commits; the action choice is
calibrated offline and needs no labels once deployed. The two
decisions can disagree: diagnosing the cause correctly does not always
identify the best fix. The chosen action defaults to the physical-layer response and departs from it
only where a fix clearly helps, so every departure is confined to a small,
auditable set.}
\label{fig:pipeline}
\end{figure}
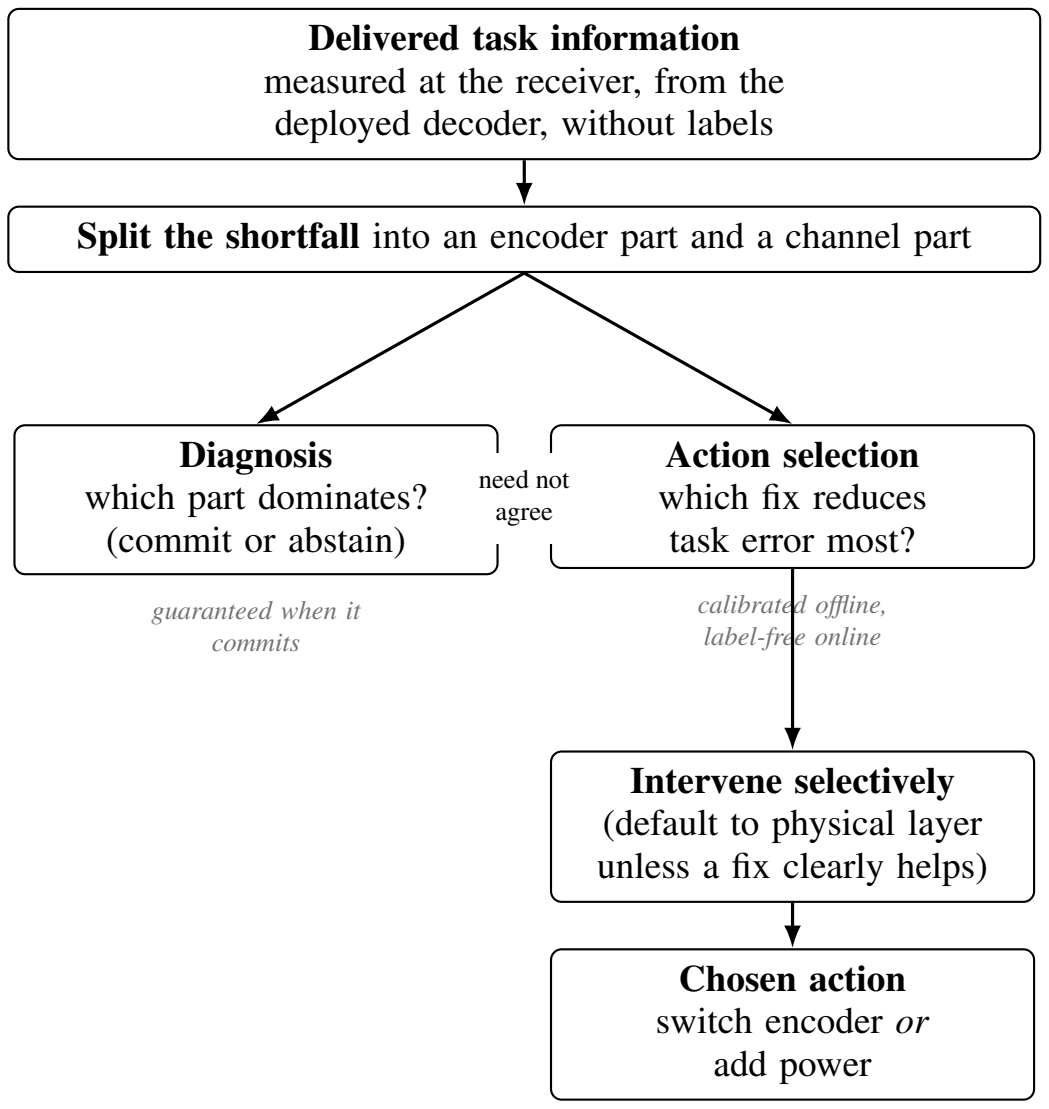

Figure~\ref{fig:pipeline} shows how the framework organises these three requirements. A single receiver-side measurement of delivered task information is partitioned into an encoder part and a channel part. The two parts drive two separate decisions: naming the responsible component, and choosing the fix that most reduces task error. The choice of fix is based on the size of each shortfall rather than on the diagnosis itself. A correct diagnosis and the best fix can therefore point in different directions; this divergence is the central concern of the work. The remainder of this section develops each stage of the framework in turn.

\subsection{System model and losses}
\label{ssub:model}
Let $Y$ be the task variable, taking values in a finite set $\Yset$ with entropy $H(Y)$, and let $X$ be the source observation. An encoder maps $X$ to a power-normalised transmitted signal $\Ztx$. A channel with known state $S\sim P_S$ transforms $\Ztx$ into the received signal $\Zrx$. The system obeys the Markov chain $Y\to X\to\Ztx\to\Zrx$, with $Y$ independent of $S$. We measure task information at two points: $\Itx=I(\Ztx;Y)$ immediately after the encoder, and $\Irx=I(\Zrx;Y\mid S)$ at the receiver, averaged over $P_S$. The data-processing inequality (DPI) then yields two non-negative losses,
\begin{equation}
\Lenc = I(X;Y)-\Itx, \qquad \Lch = \Itx-\Irx,
\label{eq:losses}
\end{equation}
the first caused by the encoder and the second by the channel. These two quantities are the coordinates on which measurement, attribution, and control all operate. We write $M=\Lch-\Lenc$ for the signed margin between them.

\begin{assumption}[Task determinacy]\label{as:determinacy}
$H(Y\mid X)=0$, so $I(X;Y)=H(Y)$.
\end{assumption}
The condition is exact when $Y$ is a deterministic function of $X$, as in class-conditional generative sources, and approximate otherwise. Under a violation, $M$ is biased downward, that is, toward appearing ``encoder-limited'', by exactly $H(Y\mid X)$.

Every guarantee stated below rests on the same notion of validity, which we fix here.

\begin{definition}[Certified]\label{def:certified}
A statement is called \emph{certified} if it is population-valid, conditional on the one-sided estimators being valid bounds on the true population information quantities defined in \eqref{eq:losses}.
\end{definition}
The property follows deterministically from the DPI once that bracketing holds. It is not a finite-sample guarantee: finite-sample coverage, where required, must be supplied by an external risk-control layer rather than by the certificate itself.

\subsection{Delivered task information and the task-error bracket}
\label{ssub:metric}
\begin{definition}[Semantic SNR]\label{def:ssnr}
We define $\mathrm{S\text{-}SNR}=\Irx/(H(Y)-\Irx)$. This quantity increases strictly with the delivered task information $\Irx$.
\end{definition}
Semantic SNR is a monotone reparameterisation of $\Irx$ (equivalently, of $\Irx/H(Y)$), so every claim made about semantic SNR applies equally to $\Irx$ itself.

\begin{theorem}[Task-error sandwich]\label{thm:sandwich}
Write $\rho$ for the semantic SNR and $K=|\Yset|$. The smallest achievable task error $\Pe$ obeys
\begin{equation}
\frac{1}{\log_2 K}\!\left(\frac{H(Y)}{1+\rho}-1\right) \le \Pe \le
\frac12\cdot\frac{H(Y)}{1+\rho},
\label{eq:sandwich}
\end{equation}
where both endpoints decrease strictly as $\rho$ grows. The gap between the two endpoints shrinks as $\rho$ grows, but only when $K>4$ (proof: supplementary material).
\end{theorem}
The bracket is obtained by combining Fano's inequality \cite{cover2006} with the Hellman--Raviv bound \cite{hellman1970}. An estimate of $\Irx$ may be substituted into these two bounds only in the direction that preserves validity: the certified upper estimate $\Ihatup\ge\Irx$ preserves the Fano lower bound, and the Barber--Agakov lower estimate \cite{barber2003} $\Ihatdn\le\Irx$ preserves the Hellman--Raviv upper bound, because each bound is decreasing in $\Irx$.

\subsection{Label-free estimation and the margin slack}
\label{ssub:estimation}
At deployment the receiver possesses neither the original source signal nor task labels. We employ a one-sided pair of estimates. A certified upper estimate, $\Ihatup=\min\{H(Y),C_{\mathrm{lat}},\mathrm{CLUB}\}$, combines the task entropy, the capacity $C_{\mathrm{lat}}$ of the latent link, and the CLUB bound \cite{cheng2020}. A lower estimate, $\Ihatdn=H(Y)-\mathrm{CE}$, is the Barber--Agakov bound computed from the decoder's own cross-entropy \cite{barber2003}. A separate temperature is fitted for each encoder, offline, by minimising negative log-likelihood \cite{guo2017}, rather than pooling the encoders under a single shared temperature. Online, the labelled cross-entropy is replaced by a predictive-information proxy $I_q(\Zrx;Y)$, computed under the decoder's own posterior. This proxy equals the true population quantity only when the decoder posterior matches the true conditional distribution, which is why the bound is population-conditional rather than exact.

Attribution also requires one-sided estimates of $\Itx$. On the transmitted (clean-path) side we reuse the Barber--Agakov construction applied to the decoder's posteriors on the transmitted signal, yielding the lower estimate
\begin{equation}
\Itxhatdn = H(Y)-\mathrm{CE}(\Ztx),
\label{eq:itx_lower}
\end{equation}
which is computable offline because the receiver's decoder can be run on $\Ztx$ directly, without the intervening channel.

For the transmitted-side upper estimate the simplest choice is the trivial bound $H(Y)$, which carries no information about the encoder at all. We instead reuse the same CLUB construction already applied on the received side, now applied to the clean-path decoder posteriors $\hat p(Y\mid\Ztx)$, which are available offline by running $\Ztx$ through the decoder with no channel noise:
\begin{equation}
\Itxhatup = \min\{H(Y),\,\mathrm{CLUB}(\hat p(Y\mid \Ztx))\}.
\label{eq:itx_tighter}
\end{equation}
This is a valid upper bound because the minimum of two valid upper bounds is itself a valid upper bound, and CLUB upper-bounds the population mutual information under the same calibration assumption already relied upon for the received side \cite[Prop.~1]{cheng2020}. By construction, \eqref{eq:itx_tighter} reduces to the trivial bound whenever the clean-path posterior is nearly deterministic, and improves on it in proportion to the residual task uncertainty that posterior retains, which is the regime in which the encoder is rate-constrained.

These four one-sided estimates bracket the signed margin. Using $I(X;Y)=H(Y)$ from Assumption~\ref{as:determinacy}, $M=2\Itx-\Irx-H(Y)$ satisfies
\begin{equation}
M\in[2\Itxhatdn-\Ihatup-H(Y),\ 2\Itxhatup-\Ihatdn-H(Y)],
\label{eq:margin_interval}
\end{equation}
an interval of half-width
\begin{equation}
\Sigma=\tfrac12[2(\Itxhatup-\Itxhatdn)+(\Ihatup-\Ihatdn)].
\label{eq:sigma}
\end{equation}
We write $\widehat M$ for the midpoint of the interval in \eqref{eq:margin_interval}; this is the observable quantity the deployed rule tests, whereas $M$ itself is not observable at run time. The factor of two on the encoder term in \eqref{eq:sigma} is the reason $\Sigma$ is dominated by looseness on the transmitted side, and therefore why the tighter bound of \eqref{eq:itx_tighter} enters the commitment test with disproportionate weight.

\subsection{Certified diagnosis}
\label{ssub:attrib}
The controller has two possible actions, $\mathcal{A}=\Ach\cup\Aenc$: spend more power, or switch to a better encoder. Which of the two the shortfall calls for is fixed by the sign of the margin.
\begin{definition}[Loss-attribution regime]\label{def:regime}
An operating point is called channel-limited if $\Lch>\Lenc$, and encoder-limited if $\Lenc>\Lch$. We write $\Delta=|M|$ for the margin between them. Ties are assigned to the channel-limited class, since spending power is the cheaper of the two fixes.
\end{definition}

The interval \eqref{eq:margin_interval} then yields a diagnosis that either commits or declines to.
\begin{theorem}[Certified diagnosis]\label{thm:certified}
Consider the rule that declares a regime only when the certified margin interval falls entirely on one side of zero, and abstains otherwise. This rule is always correct when it commits. Abstention implies $\Delta\le2\Sigma$, so whenever $\Delta>2\Sigma$ the rule is guaranteed to commit. A tighter threshold, $\Delta\le\Sigma$, describes abstention exactly only in the special case where the certified interval happens to be centred on the true margin (proof: supplementary material).
\end{theorem}
If Assumption~\ref{as:determinacy} is relaxed to allow $\epsilon=H(Y\mid X)\le\bar\epsilon$ for some known bound $\bar\epsilon$, the margin interval widens on the channel-limited side and its half-width becomes $\Sigma_\epsilon=\Sigma+\bar\epsilon/2$.
\begin{corollary}[Determinacy-robust diagnosis]\label{cor:eps}
Theorem~\ref{thm:certified} holds without change if $\Sigma$ is replaced by $\Sigma_\epsilon=\Sigma+\bar\epsilon/2$. The guarantee weakens gracefully as $\bar\epsilon$ grows, and it reduces exactly to Theorem~\ref{thm:certified} when $\bar\epsilon=0$ (proof: supplementary material).
\end{corollary}

\subsection{Limits of diagnosis-driven control}
\label{ssub:limits}
A diagnosis names the dominant loss component. Turning that name into an action requires two further conditions, and the second of them is where discrete encoder switching bites.
\begin{assumption}[Directed response]\label{as:response}
A channel-directed action changes only $\Lch$; an encoder-directed action changes only $\Lenc$.
\end{assumption}
The action set is defined to satisfy this condition: a channel-directed action is a power increase at a fixed encoder, and an encoder-directed action is a switch to a higher-rate encoder at fixed power. Actions that move both coordinates simultaneously lie outside the model developed here.
\begin{assumption}[Order consistency]\label{as:orderconsistency}
Whenever the decision rule selects class $c$, its certified reduction interval $\hat R_c$, obtained by propagating the task-error bracket of Theorem~\ref{thm:sandwich} through the action of class $c$, is not dominated by the interval $\hat R_{c'}$ of the other class.
\end{assumption}
Monotonicity of the task-error response constrains each loss coordinate individually. It does not relate the sensitivity of the response to one coordinate against the other, so Assumption~\ref{as:orderconsistency} is a separate condition rather than a consequence of monotonicity. It is a property of the system rather than of the construction, and it is least likely to hold once encoders are switched over a coarse, discrete set.

\begin{theorem}[Attribution optimality]\label{thm:attribution}
Let $W$ be the width of the task-error bracket, and let $R_{\mathrm{enc}}$ and $R_{\mathrm{ch}}$ be the true error reductions from the best action of each class. Under Assumption~\ref{as:response} and a monotone task-error response, the regret of the chosen class, $\Gamma=\max\{R_{\mathrm{enc}},R_{\mathrm{ch}}\}-R_{\mathrm{chosen}}$, satisfies $\Gamma\le4W$ under Assumption~\ref{as:orderconsistency}, and this constant cannot be improved. If the two certified reduction intervals do not overlap and the rule selects the higher one, then $\Gamma=0$ exactly (proof: supplementary material).
\end{theorem}

Theorem~\ref{thm:attribution} therefore provides a conditional baseline. It characterises the best case achievable by diagnosis-driven control when Assumption~\ref{as:orderconsistency} holds. It does not apply when this condition fails. Sec.~\ref{ssub:control} therefore develops a controller that does not rely on this assumption.

\subsection{Action-value control}
\label{ssub:control}
Real deployments switch among a handful of separately trained, discretely rate-adapted encoders that can differ in more ways than the single rate coordinate tracked by the decomposition. As a result, a simple comparison of $\Lenc$ against $\Lch$ can mislead in two distinct ways. First, the two estimated deficits can be mis-scaled relative to one another: each is individually informative about its own class, but comparing their raw magnitudes is not. Second, the sign of the estimated margin can be anti-correlated with the actual error reduction, in which case a rule that acts on this sign is worse than one that ignores it entirely. The controller of Definition~\ref{def:actionvalue} is constructed to be robust to both failure modes.

\begin{definition}[Calibrated action-value map]\label{def:actionvalue}
For each action class $c\in\{\mathrm{enc},\mathrm{ch}\}$, a map $g_c$ is fitted offline under a fixed, a priori \emph{nondecreasing} isotonic constraint \cite{barlow1972}: a larger estimated loss deficit of class $c$ is constrained to predict an error reduction no smaller than that of a smaller deficit of the same class. The direction is fixed in advance and is identical for both classes; it is not selected from the calibration data. The map sends the estimated loss deficit of class $c$, written $\widehat L_c$ and equal to the label-free estimate of $\Lenc$ for $c=\mathrm{enc}$ and of $\Lch$ for $c=\mathrm{ch}$, to the error reduction that the best action of that class actually produces. It depends only on that class's own deficit.
\end{definition}

The model is trained using a labelled calibration split but operates without labels during deployment. We refer to this approach as action-value control, which is offline-supervised and online label-free. This distinguishes it from the population-valid diagnosis in Sec.~\ref{ssub:attrib}, which provides a formal, although conditional, guarantee that action-value control does not.

The controller then selects $\hat c=\argmax_c g_c(\widehat L_c)$. Each $g_c$ is nondecreasing in its own class's deficit alone; it never takes the other class's deficit, or their difference, as input. This is why the rule remains valid even when the raw cross-class margin's sign is anti-correlated with the true reduction: the prediction of $g_c$ depends only on the magnitude of $\widehat L_c$ within class $c$, so a failure of the raw cross-class comparison cannot, by construction, alter the direction of $g_c$ itself. Where the deficit carries little exploitable signal, the fitted map flattens and the rule falls back toward whichever class is more common overall. This is a tendency rather than a guarantee: a finite-sample isotonic fit can retain several blocks even when the underlying relationship is weak.

\begin{proposition}[Isotonic consistency, informal]\label{prop:isotonic}
Let $r_c$ denote the realised task-error reduction produced by the best action of class $c$, and let $\rho_c(x)=\mathbb E[r_c\mid\widehat L_c=x]$ be a bounded function. As the size of the calibration split grows, the fitted map $g_c$ converges, at every point of continuity, to the nondecreasing isotonic projection of $\rho_c$; if $\rho_c$ is itself already nondecreasing, $g_c$ converges exactly to $\rho_c$. If $\rho_c$ is not nondecreasing, the projection is the closest nondecreasing approximation to it in the least-squares sense under the empirical measure, not $\rho_c$ itself \cite{barlow1972}.
\end{proposition}
This is the standard consistency result for isotonic regression, stated here for Definition~\ref{def:actionvalue}. It is a known result rather than a new theoretical contribution. It also has three important limitations. First, it provides no convergence rate for the calibration sizes used in our experiments. Second, it does not require $\rho_c$ to be nondecreasing. Third, it does not establish that $\argmax_c g_c$ is consistent across the two maps. It only establishes consistency for each map separately. Therefore, the action-selection behaviour of the deployed rule is evaluated empirically rather than guaranteed by this proposition.

 The calibrated rule is not used on its own. We combine it with a simple default policy based only on the instantaneous channel quality. The controller follows this default unless the calibrated action values disagree with it by more than a small margin. Thus, the controller changes the default only at a limited set of operating points. These intervention points can be counted and audited directly. When the deficit contains little exploitable information, the fitted action-value maps become nearly flat. The disagreement margin is then rarely exceeded, resulting in a small override set.

\subsection{Two kinds of guarantee, and how they compose}
\label{ssub:certvsdeploy}
The framework now contains two decision rules, and they carry different kinds of guarantee. Keeping the two apart matters, because they are not interchangeable.

The certified diagnosis of Theorem~\ref{thm:certified} is population-valid diagnosis conditional on valid one-sided bounds: whenever the one-sided estimators of $\Itx$ and $\Irx$ are valid bounds on their true population values, every commitment the rule makes is correct, by the deterministic argument of Definition~\ref{def:certified}. The action-value controller of Definition~\ref{def:actionvalue} is a different object. Its isotonic maps are fitted on a labelled calibration split, but the deployed rule evaluates them at label-free estimated loss deficits online and never abstains. Because it does not inherit the certified rule's bracketing argument, it carries no formal regret guarantee of its own: Proposition~\ref{prop:isotonic} gives asymptotic consistency for each map individually and stops short of a guarantee for the comparison between them.

Under the trivial bound $H(Y)$, the certified diagnosis abstains whenever the margin interval straddles zero. The tighter bound $\Itxhatup$ narrows that interval and therefore widens the region in which the rule commits. This affects only how often the diagnosis commits, not whether the action-value controller's selections are guaranteed. Neither $\Sigma$ nor Theorems~\ref{thm:sandwich}--\ref{thm:attribution} accounts for the finite-sample randomness of the underlying estimators.

Finite-sample reliability, where an application demands it, therefore has to come from outside the framework. It composes without interference.
\begin{proposition}[Coverage-preserving composition]\label{prop:coverage}
Suppose a distribution-free risk-control layer \cite{angelopoulos2023}
certifies a task-error target at confidence $1-\alpha$ uniformly over
an admissible set of configurations $\mathcal{C}$. Then any controller
whose selected action remains in $\mathcal{C}$ preserves the same
$1-\alpha$ guarantee, regardless of how it selects among those
certified configurations (proof: supplementary material).
\end{proposition}
In other words, attribution can affect only which feasible action is chosen; it does not affect whether the coverage target is met, provided every selected action remains within the certified set.

\subsection{Algorithms and complexity}
\label{ssub:algo}
Algorithm~\ref{alg:offline} performs the offline steps. It fits a temperature for each encoder, computes the tighter bound $\Itxhatup$, computes the slack $\Sigma$, and fits the two isotonic maps. Algorithm~\ref{alg:online} performs the corresponding online steps. It first computes label-free loss estimates, checks them against the certified rule in Theorem~\ref{thm:certified}, and then applies the calibrated action together with the channel-quality default.

Two points about the online procedure require clarification. First, the commitment test compares the estimated margin midpoint $\widehat M$ in \eqref{eq:margin_interval} with $\Sigma$. This is exactly the rule analysed in Theorem~\ref{thm:certified}. The certified interval excludes zero when $|\widehat M|>\Sigma$. This should not be confused with the condition $\Delta>2\Sigma$ that also appears in the theorem. There, $\Delta=|M|$ is the true margin and is unobservable at run time. The condition $\Delta>2\Sigma$ is only a sufficient condition for the implemented test to fire. It is not the test itself.

Second, the external risk-control layer in Proposition~\ref{prop:coverage} restricts the available action set. This restriction does not come from the diagnostic certificate. Theorem~\ref{thm:certified} certifies the declared regime but does not guarantee that any particular action achieves a target task error.

The online computational cost is small. It scales linearly with the alphabet size for computing the information estimates and with the number of actions for evaluating the constant-time action-value maps. This cost is small relative to the decoding operation already performed by the receiver. The offline cost scales linearly with the calibration-set size. No additional neural network is trained in the pipeline.

\begin{algorithm}[t]
\caption{Offline characterisation (labelled calibration split)}
\label{alg:offline}
\begin{algorithmic}[1]
\State fit per-encoder calibration temperatures by minimising NLL
\State compute clean-path posteriors and form $\Itxhatup$ via~\eqref{eq:itx_tighter}
\State compute $\Ihatup,\Ihatdn$; derive margin slack $\Sigma$
\State fit isotonic action-value maps $g_{\mathrm{enc}},g_{\mathrm{ch}}$
\State \textbf{output:} temperature, $\Itxhatup$, $\Sigma$, $g_{\mathrm{enc}},g_{\mathrm{ch}}$
\end{algorithmic}
\end{algorithm}
\begin{algorithm}[t]
\caption{Online cause-attributed control (label-free)}
\label{alg:online}
\begin{algorithmic}[1]
\State estimate delivered information from calibrated decoder posteriors
\State form loss estimates $\widehat{\Lenc},\widehat{\Lch}$; predict $v_c=g_c(\widehat{L}_c)$; $\hat c=\argmax_c v_c$
\If{$|\widehat M|>\Sigma$} record certified regime (Thm.~\ref{thm:certified}) \EndIf
\If{$\hat c$ differs from channel-quality default by margin $>0$}
    \State restrict $\mathcal{A}$ to actions permitted by the external risk-control layer; apply class $\hat c$
\Else\ \State apply the permitted channel-quality default \EndIf
\end{algorithmic}
\end{algorithm}

\section{Result Analysis}
\label{sec:results}
\subsection{Experimental protocol}
\label{ssub:setup}
We evaluate on four settings, summarised in Table~\ref{tab:setup}. They differ in task difficulty, image resolution, and modality, and together they permit separation of questions about estimation from questions about control.

The controlled model is a $K{=}10$-class Gaussian-mixture channel model, chosen because $\Itx$, $\Irx$, and the Bayes error all possess closed-form expressions. It satisfies Assumption~\ref{as:determinacy} exactly by construction. Class means are placed close together relative to the encoder noise, which elevates absolute error (above one half at many operating points). The model is a stress test for the bracket and the attribution machinery; it is not intended as a difficulty-ranked benchmark comparable to the learned datasets. Table~\ref{tab:main} should therefore be read column by column, not compared across columns.

The three DeepJSCC systems share one architecture: two stride-2 convolutions to width $w,2w$, a linear map to $k$ complex symbols under a unit power constraint, and a two-layer decoder. All three are trained with Adam ($10^{-3}$ learning rate, batch size 256, cross-entropy loss, training SNR of 6\,dB) and evaluated from $-6$ to $+12$\,dB in 3\,dB steps. Calibration and evaluation always use disjoint halves of the test split.

\begin{table}[t]
\caption{Experimental settings.}
\label{tab:setup}
\centering\scriptsize\setlength{\tabcolsep}{3pt}\renewcommand{\arraystretch}{1.05}
\begin{tabular}{lccccc}
\toprule
Setting & $K$ & Rate pts.\ $k$ & Width & Grid & Seeds \\
\midrule
Controlled & 10 & --- & --- & $6{\times}8{=}48$ & closed-form \\
MNIST (easy) & 10 & $2,4,8,16,32$ & 32 & $4{\times}7{=}28$ & 20 \\
CIFAR-10 (hard) & 10 & $1,2,3,4,6$ & 14 & $4{\times}7{=}28$ & 20 \\
STL-10 (aggr.) & 10 & $2,3,4,6,8$ & 16 & $4{\times}7{=}28$ & 10 \\
Wine (pilot) & 3 & --- & --- & $3{\times}5{=}15$ & 5 \\
\bottomrule
\end{tabular}
\end{table}

At every operating point the controller may take one action: spend more power, or switch to a higher-rate encoder. The oracle tries both and retains whichever reduces the held-out task error more. Action-selection accuracy is simply the frequency with which a policy agrees with the oracle's choice. On the controlled model the oracle is exact, because the loss-attribution regime is available in closed form. On the three learned systems the oracle is an empirical construct: both actions are executed and the resulting held-out errors are compared. Because a single noisy comparison can be unstable when the true gap between the two actions is small, five independent noise draws are averaged per action before the choice is made. We call this the five-draw stabilised oracle and use it for every learned-system number reported. A separate check (varying the number of draws from 1 to 50, Sec.~\ref{ssub:res_tighter}) confirms that the proxy is stable at the fixed training seed employed.

Every cause-agnostic baseline against which we compare---channel SNR, decoder confidence, a CIBA-style feature-importance rule \cite{ciba}, a hard-decision BER rule, a majority-class rule, and, for the metric comparison only, an uncalibrated STII-style proxy \cite{stii}---uses only information available at the current operating point, with no look-ahead at candidate actions. Realised errors are compared by paired Wilcoxon signed-rank tests and a paired bootstrap with $B{=}3000$ resamples. All reported means carry $95\%$ bootstrap confidence intervals, and binomial quantities carry Clopper--Pearson intervals. Action-selection accuracies in the tables are point estimates only, because the corresponding confidence intervals at these grid sizes are wide (roughly $\pm0.1$); statistical claims therefore rest on paired tests over matched operating points, reported with exact $p$-values.

\subsection{Certified bound, predictive-information proxy, and metric comparison}
\label{ssub:res_sandwich}
\begin{figure}[t]
\centering
\figorbox{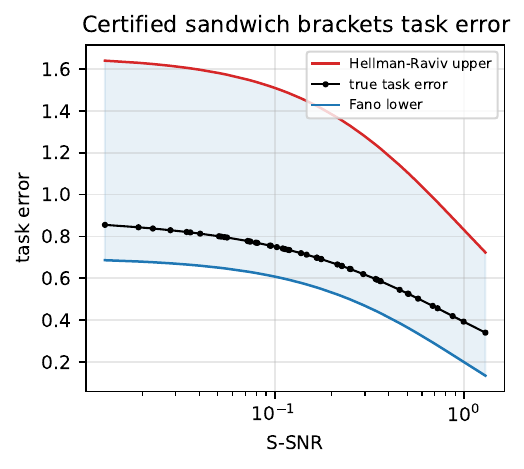}{0.85\columnwidth}{Fano lower / Hellman--Raviv upper endpoints bracketing true task error vs.\ S-SNR.}
\caption{Certified sandwich (Theorem~\ref{thm:sandwich}): the bracket contains true task error at every point and tightens with semantic SNR.}
\label{fig:sandwich}
\end{figure}

Figure~\ref{fig:sandwich} plots the certified bracket of Theorem~\ref{thm:sandwich} against semantic SNR. The bracket contains the true task error at $100\%$ of the controlled model's 48 operating points (Clopper--Pearson interval $[0.926,1.000]$) and tightens smoothly as semantic SNR increases. Substitution of the label-free estimate for the true delivered information still preserves the certified lower endpoint at $100\%$ of points.

The decoder-derived predictive-information proxy is already strongly rank-faithful before calibration (Pearson $r{=}0.98$), and this rises to $r{=}1.00$ after temperature calibration. Before calibration, points are shifted consistently off the diagonal, indicating that the proxy is biased rather than merely noisy. Calibration therefore removes this bias without altering the ranking of operating points. An uncalibrated proxy is consequently safe for ranking points but must not be substituted directly into the bound endpoints. Separately, the received-side CLUB bound still leaves substantial slack in the regimes studied: it yields a mean margin slack of $2.08$ bits, an improvement on the $2.31$ bits of the simpler capacity-based estimate but far too wide for the rule to commit anywhere. This residual slack is what motivates the tighter transmitted-side bound of Sec.~\ref{ssub:estimation}.

Table~\ref{tab:metric} compares several allocation signals under one fixed decision rule. The calibrated semantic-SNR decomposition selects the oracle-best action class $87.5\%$ of the time, against $85.4\%$ for the uncalibrated STII-style proxy and $75.0\%$ for channel SNR alone. Decoder confidence, the CIBA-style rule, and the BER rule all sit at only $29.2\%$, because each can detect that a link is failing without identifying the cause. Because the uncalibrated proxy performs almost as well as the calibrated one, most of the gain over the simple physical-layer proxies arises from the decomposition itself rather than from calibration.

\begin{table}[t]
\caption{Allocation-signal comparison, fixed raw point-diagnosis rule (48 pts). ``Calibrated/uncalibrated'' refers to the information estimator, not the action-value rule.}
\label{tab:metric}
\centering\footnotesize\setlength{\tabcolsep}{3pt}\renewcommand{\arraystretch}{1.05}
\begin{tabular}{p{2.3cm}cp{2.6cm}}
\toprule
\textbf{Signal} & \textbf{Acc.} & \textbf{Capability} \\
\midrule
S-SNR decomp.\ (ours) & 0.875 & Both components; calibrated. \\
STII proxy \cite{stii} & 0.854 & Both components; uncalibrated. \\
Channel SNR & 0.750 & Channel only; blind to encoder. \\
CIBA / BER / Conf.\ & 0.292 & Detect failure, not cause. \\
Majority baseline & 0.708 & --- \\
\bottomrule
\end{tabular}
\end{table}

\subsection{Regime separation and sensitivity to source-task ambiguity}
\label{ssub:res_attrib}
\begin{figure}[t]
\centering
\figorbox{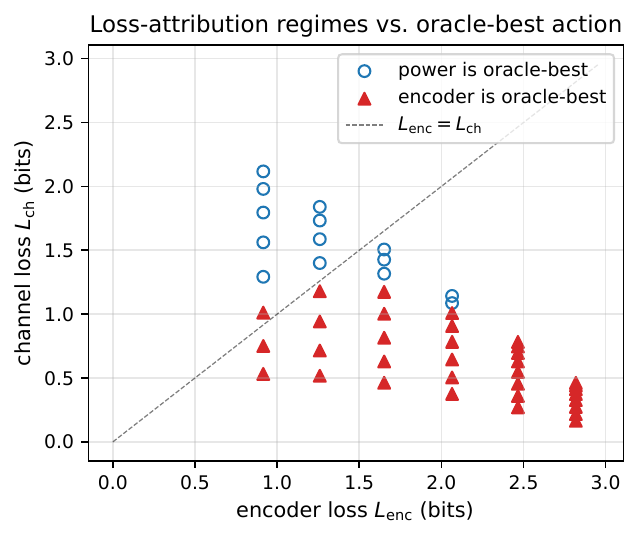}{0.8\columnwidth}{Channel vs.\ encoder loss per point, marked by oracle-best action.}
\caption{Loss-attribution regimes are separable: equal-error points fall on both sides of $\Lenc=\Lch$.}
\label{fig:decomp}
\end{figure}

Figure~\ref{fig:decomp} shows the two loss components at every operating point. Points with nearly identical task error can fall on either side of the $\Lenc{=}\Lch$ diagonal; in fact, three of the five equal-error bins on the controlled model contain both regimes. In other words, two points with the same task error can genuinely require opposite fixes. This is the core premise of the framework.

Under the trivial bound, the slack $\Sigma$ on the controlled model has mean $2.31$ bits, standard deviation $0.51$, and range $[1.17,3.07]$ bits. Because the smallest value of $\Sigma$ still exceeds the largest margin anywhere on the grid, the certified rule abstains at every single point under this bound. The situation changes once the tighter bound $\Itxhatup$ is substituted (Sec.~\ref{ssub:res_tighter}).

We also test the sensitivity of the diagnosis to a violation of Assumption~\ref{as:determinacy}. A hypothetical residual entropy $H(Y\mid X)=\bar\epsilon$ is swept and the number of declaration flips, together with the number of points on which the rule still commits, is recorded at the two endpoints $\bar\epsilon=0$ and $\bar\epsilon=0.50$ bits. Table~\ref{tab:hyx_combined} reports the results for all four backends under both the trivial and the tighter bound.

On the controlled model, where ground truth permits a direct check of the sweep, the bias is small and grows slowly: even at the extreme value $\bar\epsilon=0.50$ bits (larger than any plausible labelling ambiguity on a real dataset), only 8 of 48 declarations flip. CIFAR-10 and STL-10 are at least this robust; STL-10 exhibits zero flips over the entire sweep. MNIST is the exception. Because the tighter bound does not engage on MNIST at all (Sec.~\ref{ssub:res_tighter}), the underlying one-sided estimator bias interacts with MNIST's mostly channel-limited grid in a way that pushes many near-zero margins past the commit threshold: full-width commits more than double, from $9/28$ to $22/28$, and half of all declarations flip at $\bar\epsilon=0.50$. This sensitivity is specific to MNIST and appears only at residual-entropy values well beyond any plausible amount of labelling ambiguity on a standard benchmark.

\begin{table}[t]
\caption{$H(Y\mid X)$ sensitivity at the two swept endpoints ($\bar\epsilon=0$ and $\bar\epsilon=0.50$ bits): declaration flips and full-width certified commits, under the trivial bound ($\Itx^\uparrow{=}H(Y)$) and the tighter bound $\Itxhatup$ (Sec.~\ref{ssub:estimation}), all backends. Flip counts are identical under both bounds and are therefore reported once.}
\label{tab:hyx_combined}
\centering\scriptsize\setlength{\tabcolsep}{4pt}\renewcommand{\arraystretch}{1.05}
\begin{tabular}{lcc cc}
\toprule
& \multicolumn{2}{c}{Flips (of $n$)} & \multicolumn{2}{c}{Full-width commits, trivial $\to$ tighter} \\
Backend & $\bar\epsilon{=}0$ & $0.50$ & $\bar\epsilon{=}0$ & $0.50$ \\
\midrule
Controlled ($n{=}48$) & 0 & 8 & $0\to16$ & $0\to14$ \\
MNIST ($n{=}28$) & 0 & 14 & $9\to9$ & $9\to22$ \\
CIFAR-10 ($n{=}28$) & 0 & 2 & $0\to26$ & $0\to22$ \\
STL-10 ($n{=}28$) & 0 & 0 & $0\to28$ & $0\to28$ \\
\bottomrule
\end{tabular}
\end{table}

For the controlled model specifically, the half-width certified rule (the more permissive threshold $\Delta>\tfrac12\Sigma$), computed under the trivial bound, can also be checked. Half-width commits fall from $24/48$ at $\bar\epsilon=0$ to $18/48$ at $\bar\epsilon=0.50$, a modest drop consistent with the full-width commits in Table~\ref{tab:hyx_combined}.

\subsection{Finite-sample uncertainty on $\Sigma$}
\label{ssub:res_finitesample}
Finite-sample uncertainty of $\Sigma$ itself is estimated by a bootstrap over the calibration split (Sec.~\ref{ssub:certvsdeploy}). On the controlled model ($B{=}100$ resamples per point), the resulting $95\%$ confidence interval on $\Sigma$ has a mean relative width of $6.4\%$ and a maximum of $14.4\%$. On CIFAR-10 and STL-10 ($B{=}500$ resamples), the mean relative widths are $3.9\%$ and $3.3\%$ (maxima $5.4\%$ and $4.3\%$), narrower as expected because these calibration splits are larger. In every case measured, the finite-sample uncertainty on $\Sigma$ is small compared with $\Sigma$ itself, which bounds how large the gap can practically be between the population-conditional guarantee and a fully finite-sample-valid one.

\subsection{The tighter transmitted-side bound, structural mismatch, and the certified layer}
\label{ssub:res_tighter}
On the controlled model, where $\Itx$ has a closed form, the tighter bound $\Itxhatup\ge\Itx$ holds at all 48 points, with a minimum slack of $0.51$ bits and a mean slack of $2.35$ bits. This directly validates \eqref{eq:itx_tighter}. Table~\ref{tab:tighter} reports the mean slack $\Sigma$ and the resulting commit counts under both bounds. The tighter bound yields a large improvement precisely where the clean-path decoder still carries real task-relevant uncertainty, that is, where the encoder is genuinely rate-constrained: on CIFAR-10, $\Sigma$ drops from $2.21$ to $0.58$ and full-width commits rise from 0 to 26 out of 28; on STL-10, $\Sigma$ drops from $2.44$ to $0.27$ and commits rise from 0 to 28 out of 28. On MNIST, where the clean-path decoder is nearly deterministic and CLUB exceeds $H(Y)$ at every point, the tighter bound yields no improvement at all.

\begin{table}[t]
\caption{Mean slack $\Sigma$, full/half-width commits, and committed-point accuracy against the decomposition's dominant loss component, trivial versus tighter bound.}
\label{tab:tighter}
\centering\scriptsize\setlength{\tabcolsep}{2.5pt}\renewcommand{\arraystretch}{1.05}
\begin{tabular}{lcccc c cccc}
\toprule
& \multicolumn{4}{c}{Trivial $H(Y)$} & & \multicolumn{4}{c}{Tighter $\Itxhatup$} \\
\cmidrule(lr){2-5}\cmidrule(lr){7-10}
Backend & $\bar\Sigma$ & Full & Half & Acc. & & $\bar\Sigma$ & Full & Half & Acc. \\
\midrule
Controlled & 2.31 & 0/48 & 24/48 & --- & & 1.60 & 16/48 & 25/48 & 1.00 \\
CIFAR-10   & 2.21 & 0/28 & 23/28 & --- & & 0.58 & 26/28 & 27/28 & 1.00 \\
STL-10     & 2.44 & 0/28 & 27/28 & --- & & 0.27 & 28/28 & 28/28 & 0.80 \\
MNIST      & 0.37 & 9/28 & 21/28 & 1.00 & & 0.37 & 9/28 & 21/28 & 1.00 \\
\bottomrule
\end{tabular}
\end{table}

``Committed-point accuracy'' is scored against the decomposition's own loss ordering, which is what Theorem~\ref{thm:certified} certifies, and not against the oracle-best action. These are two different quantities; the gap between them is measured next. STL-10's $0.80$ committed-point accuracy reflects imperfect generalisation of the loss ordering from the calibration split to the test split, not a violation of the bound. The ordering is perfect on CIFAR-10, and STL-10 has the smallest per-point margins of any system.

\emph{Structural versus estimation mismatch.} Each system's total disagreement between the raw rule and the oracle is partitioned into two parts. The structural part measures how often the decomposition's own loss ordering already disagrees with the oracle-best action's class; this is a ceiling for any policy that acts on the loss ordering alone, however accurately that ordering is estimated. The estimation part measures how often the calibration-split diagnosis fails to recover that loss ordering, but only on the subset of points where the ordering was already correct. Table~\ref{tab:structural} reports both parts, computed at a single seed. All three learned systems exhibit substantial structural mismatch, ranging from $36\%$ on CIFAR-10 to $54\%$ on STL-10, and even the controlled model---where the encoder-directed action can vary continuously---shows a non-trivial $13\%$. These counts are the direct empirical measurement of how often Assumption~\ref{as:orderconsistency} fails, and therefore of how far the conditional baseline of Theorem~\ref{thm:attribution} is from describing these systems.

A separate check confirms that structural mismatch is not merely an artefact of oracle noise. The number of oracle noise-averaging draws was swept from 1 to 50; the structural-mismatch count did not change on either learned system, no operating point disagreed with a more heavily averaged 31-draw reference oracle, and the mismatched points were decided by comfortable margins (mean gap of approximately $2\%$ in realised error between the two actions) rather than by near-ties.

\begin{table}[t]
\caption{Structural mismatch (loss ordering disagrees with oracle-best action) versus estimation mismatch (diagnosis disagrees with the loss ordering, on the structurally agreeing subset), single seed.}
\label{tab:structural}
\centering\scriptsize\setlength{\tabcolsep}{3pt}\renewcommand{\arraystretch}{1.05}
\begin{tabular}{lcccc}
\toprule
Backend & $n$ & Structural & Estimation & Raw/predicted \\
\midrule
Controlled & 48 & $6/48$ (0.13) & $0/42$ (0.00) & 0.125/0.125 \\
MNIST      & 28 & $13/28$ (0.46) & $1/15$ (0.07) & 0.429/0.500 \\
CIFAR-10   & 28 & $10/28$ (0.36) & $2/18$ (0.11) & 0.429/0.429 \\
STL-10     & 28 & $15/28$ (0.54) & $6/13$ (0.46) & 0.750/0.750 \\
\bottomrule
\end{tabular}
\end{table}

On MNIST, on 2 of its 13 structurally mismatched points, the calibration-split diagnosis happens to deviate from the loss ordering in exactly the direction that still recovers the oracle's action---a coincidental cancellation, not a genuine resolution of cause. This is why MNIST's raw-rule accuracy in Table~\ref{tab:main} appears comparatively strong: it reflects estimator error landing on the favourable side by chance rather than a genuinely more reliable loss ordering. CIFAR-10 and STL-10 show no such gap.

On STL-10, a loss-ordering estimator that always recovered the dominant component correctly could reach at most $1-0.536=0.464$ action-selection accuracy on this grid, because $54\%$ of points are structurally mismatched. This value is a ceiling for diagnosis-driven loss-ordering policies, not for action prediction in general: the deployed controller reaches $0.857$ (Table~\ref{tab:main}) because the isotonic maps use the continuous magnitude of the loss deficit rather than only its binary ordering.

Table~\ref{tab:certified_oracle} scores the certified layer, under the tighter bound, against both references on the same committed set. STL-10 is the clearest case: it commits on every point ($28/28$), correctly names the dominant loss component $80\%$ of the time, yet that component matches the oracle-best action only $25\%$ of the time, consistent with the $0.46$ loss-ordering ceiling of Table~\ref{tab:structural}. The population-valid diagnosis is accurate about cause. The oracle-best action is a genuinely separate quantity, one that the action-value controller targets directly instead.

\begin{table}[t]
\caption{Certified layer accuracy against the loss ordering versus against the oracle-best action, identical committed set (tighter bound). Ceiling $=1-$structural-mismatch (Table~\ref{tab:structural}), computed over the full operating grid; it bounds loss-ordering policies applied grid-wide. Accuracy on a committed subset may exceed this value, since abstention removes points selectively.}
\label{tab:certified_oracle}
\centering\scriptsize\setlength{\tabcolsep}{3pt}\renewcommand{\arraystretch}{1.05}
\begin{tabular}{lcccc}
\toprule
Backend & Committed & vs.\ ordering & vs.\ oracle & Ceiling \\
\midrule
Controlled & 16/48 (0.33) & 1.00 & 1.00 & 0.88 \\
MNIST      & 9/28 (0.32) & 1.00 & 0.70 & 0.54 \\
CIFAR-10   & 26/28 (0.93) & 1.00 & 0.62 & 0.64 \\
STL-10     & 28/28 (1.00) & 0.80 & 0.25 & 0.46 \\
\bottomrule
\end{tabular}
\end{table}

\subsection{Cause-attributed control: main result}
\label{ssub:res_control}
\begin{figure}[t]
\centering
\figorbox{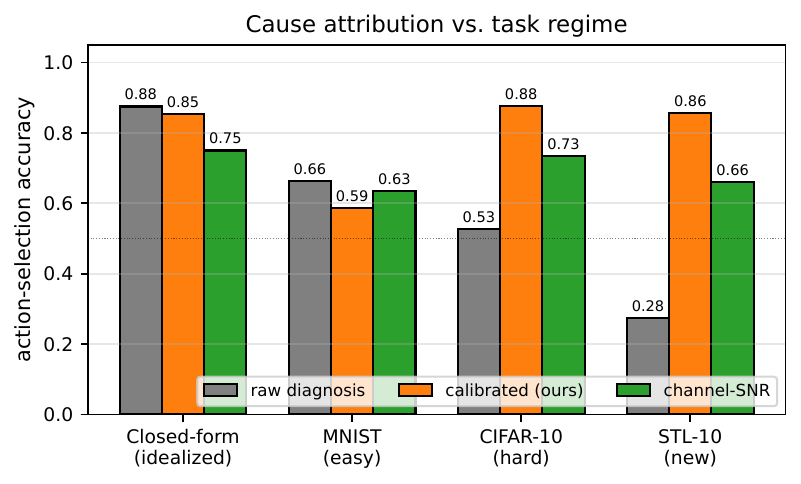}{0.92\columnwidth}{Action-selection accuracy: raw versus calibrated versus channel-SNR across four regimes.}
\caption{Calibration is decisive on CIFAR-10/STL-10, neutral on the controlled model, slightly costly on MNIST.}
\label{fig:three}
\end{figure}

Table~\ref{tab:main} and Figure~\ref{fig:three} present the central result. Each system is examined in turn.

On the controlled model, both raw and calibrated diagnosis outperform channel SNR ($0.875$ and $0.854$ versus $0.750$).

On MNIST the encoder is rarely the bottleneck, and calibration does not improve action selection. The calibrated rule reaches $0.588{\pm}0.085$, against $0.634{\pm}0.081$ for channel SNR (Wilcoxon $p{=}0.0010$, $d_z{=}{-}1.26$) and $0.664{\pm}0.076$ for raw diagnosis. The fused controller therefore provides no benefit over the channel-SNR default and incurs a small absolute realised-error penalty of $0.0006$. It departs from that default on $7.5\%$ of operating points, against $25.0\%$ on CIFAR-10 and $27.5\%$ on STL-10 (Table~\ref{tab:main}). A Gaussian-process (GP) surrogate baseline \cite{gpcr} leads here at $0.804{\pm}0.107$: MNIST possesses a smooth, two-dimensional performance surface that a learned interpolator can exploit without resolving cause.

On CIFAR-10 the raw rule actually falls below its own majority baseline ($0.527$ versus $0.536$), yet isotonic calibration on exactly the same underlying information reaches $0.877{\pm}0.046$ (versus channel SNR's $0.734{\pm}0.051$; $p{=}8.9{\times}10^{-5}$, $d_z{=}{+}2.37$ on action selection and $d_z{=}{-}3.44$ on realised error, with all 20 seeds favouring calibration). The underlying deficit was informative all along; the raw magnitude comparison simply mis-scaled the two classes relative to each other. This is a direct failure of Assumption~\ref{as:orderconsistency}, and calibration corrects it.

On STL-10 the raw diagnosis is not merely mis-scaled; it is actively anti-correlated with the true reduction. Its accuracy is $0.275{\pm}0.088$, below the $0.500$ level expected from random guessing. A per-seed check finds that 21 of 28 points disagree with the oracle---a pattern consistent with a deterministic rule that has the wrong sign, not with random noise. The isotonic map is fitted against the measured calibration delta for its own class alone, and never against the raw cross-class margin, so an inversion of that margin's sign does not enter the fit. It recovers accuracy to $0.857{\pm}0.082$, statistically indistinguishable from the GP surrogate ($0.871{\pm}0.063$, $p{=}0.753$) and far above channel SNR ($0.661{\pm}0.072$; $p{=}0.0051$, $d_z{=}{+}2.73$ on action selection and $d_z{=}{-}2.50$ on error, with all 10 seeds favouring calibration).

Taken together, the three systems exhibit three distinct mechanisms behind the same isotonic map. On MNIST the deficit carries too little signal to calibrate. On CIFAR-10 an informative deficit is mis-scaled, and calibration rescales it. On STL-10 an informative deficit is sign-inverted, and calibration corrects the sign. This is consistent with Definition~\ref{def:actionvalue}: $g_{\mathrm{enc}}$ and $g_{\mathrm{ch}}$ are each nondecreasing functions of their own class's deficit only, so neither has access to the raw cross-class margin whose sign is inverted. The within-class relationship each map estimates can be nondecreasing and correctly recovered even when the cross-class raw-margin comparison points the wrong way. A measurable consequence follows. Because $g_c$ is constrained nondecreasing, a within-class relationship that was genuinely decreasing would be projected onto a near-constant map, and the rule would degenerate to the majority class. The fitted encoder-side map on STL-10 instead has the largest coefficient of variation of any system, CV$(g_{\mathrm{enc}}){=}1.11$ (Table~\ref{tab:gate}), so the within-class relationship it estimates is not merely non-degenerate but the most structured of the three learned systems. The inversion is therefore located in the cross-class comparison, not in either within-class relationship. The effect reverses direction between MNIST on one side and CIFAR-10/STL-10 on the other: attribution helps exactly where the encoder can be the bottleneck, and is unnecessary, even slightly costly, where it cannot. This reversal is the principal empirical evidence for the separation claimed in Sec.~\ref{sec:intro}.

\emph{Comparison with a learned surrogate.} The closest prior action-value method \cite{gpcr} is a Gaussian-process surrogate that maps observable operating-point features directly to realised error, fitted on the same calibration split used for the isotonic maps. This surrogate leads on raw action-selection accuracy on three of the four systems (controlled $0.938$; MNIST $0.804$; CIFAR-10 $0.932$) and ties with the present method on STL-10 ($0.871$ versus $0.857$). Its realised error is statistically indistinguishable from that of the fused controller on the controlled model, CIFAR-10, and STL-10, and significantly lower on MNIST ($p{=}8.9{\times}10^{-5}$, $d_z{=}{+}2.06$, with all 20 seeds favouring the surrogate). It performs best precisely where the underlying performance surface is smooth and calibration data is generous. On the controlled model, CIFAR-10, and STL-10 the isotonic maps match or approach this accuracy while remaining one-dimensional and monotone, and, as shown in Sec.~\ref{ssub:res_dataefficiency}, require substantially less calibration data. On MNIST, where the deficit carries little signal for either method, the GP surrogate leads by a wide margin. Unlike the separate certified-diagnosis layer, the deployed action-value rule does not abstain; it commits to an action at every operating point. The contribution is therefore a calibration-efficient, interpretable action-value learner that is competitive with a much less constrained interpolator, not a claim of higher raw accuracy.

\begin{table*}[t]
\caption{Cause-attributed control across four regimes. Learned-system values are mean$\pm$s.d.\ over 20 seeds (MNIST, CIFAR-10) / 10 seeds (STL-10), five-draw stabilised oracle.}
\label{tab:main}
\centering\scriptsize\setlength{\tabcolsep}{2.2pt}\renewcommand{\arraystretch}{1.05}
\begin{tabular}{p{2.15cm}cccc}
\toprule
\textbf{Metric} & \textbf{Closed form} & \textbf{MNIST} & \textbf{CIFAR-10} & \textbf{STL-10} \\
\midrule
\multicolumn{5}{l}{\emph{Realised task error (lower better)}}\\
Channel SNR & 0.6291 & $0.0940\pm0.0014$ & $0.5732\pm0.0035$ & $0.6198\pm0.0045$ \\
Fused (ours) & 0.6280 & $0.0946\pm0.0014$ & $0.5674\pm0.0037$ & $0.6141\pm0.0055$ \\
GP surrogate \cite{gpcr} & 0.6277 & $0.0922\pm0.0017$ & $0.5665\pm0.0037$ & $0.6138\pm0.0051$ \\
Oracle & 0.6270 & --- & --- & --- \\
\midrule
Oracle gain captured & 0.983 & $0.959\pm0.014$ & $0.981\pm0.011$ & $0.964\pm0.025$ \\
\midrule
\multicolumn{5}{l}{\emph{Action-selection accuracy (secondary)}}\\
Raw diagnosis & 0.875 & $0.664\pm0.076$ & $0.527\pm0.055$ & $0.275\pm0.088$ \\
Fused/calibrated (ours) & 0.854 & $0.588\pm0.085$ & $0.877\pm0.046$ & $0.857\pm0.082$ \\
Channel SNR & 0.750 & $0.634\pm0.081$ & $0.734\pm0.051$ & $0.661\pm0.072$ \\
GP surrogate \cite{gpcr} & 0.938 & $0.804\pm0.107$ & $0.932\pm0.050$ & $0.871\pm0.063$ \\
Majority & 0.708 & 0.500 & 0.536 & 0.500 \\
\midrule
SNR overrides & 0.354 & $0.075\pm0.011$ & $0.250\pm0.031$ & $0.275\pm0.065$ \\
Override harm rate & --- & 0.810 & 0.214 & 0.143 \\
\bottomrule
\end{tabular}
\end{table*}

\subsection{Policy value, override audit, and the CV$(g_{\mathrm{enc}})$ gate}
\label{ssub:res_policy}
Action-selection accuracy treats every wrong choice identically, whether the two actions differ substantially or almost not at all. The primary metric is therefore realised task error, and specifically the fraction of the oracle's achievable gain that a policy actually captures: $G_{\mathrm{cap}}=\mathbb E_x[(E_{\mathrm{base}}-E_{\mathrm{policy}})/(E_{\mathrm{base}}-E_{\mathrm{oracle}})]$, where $E_{\mathrm{base}}$ is the realised error before any action is applied. The fused controller captures $96$--$98\%$ of the achievable gain across all four main settings (Table~\ref{tab:main}), and $91.3\%$ on the Wine pilot (Sec.~\ref{ssub:res_pilots}). It outperforms channel SNR significantly on CIFAR-10 ($p{=}8.9{\times}10^{-5}$) and STL-10 ($p{=}0.0051$). On MNIST it trails channel SNR by an absolute gap of $0.0006$; the gap is small relative to the per-seed standard deviation of either method, but it is consistent across seeds and statistically detectable ($p{=}1.0{\times}10^{-4}$, $d_z{=}{+}1.81$, 19 of 20 seeds favouring channel SNR). This is precisely the behaviour expected in a regime where the encoder is rarely the bottleneck.

The override fraction in Table~\ref{tab:main} measures how often selective intervention actually engages, and the benefit tracks that fraction closely. Auditing every override against the SNR default yields harm rates of $0.810$ on MNIST (42 overrides across 20 seeds, mean error increase $0.0086$), $0.214$ on CIFAR-10 (140 overrides across 20 seeds, mean reduction $0.0231$), and $0.143$ on STL-10 (77 overrides across 10 seeds, mean reduction $0.0209$)---the same ordering as the realised-error effect. On MNIST a majority of overrides are harmful; there are 42 of them in total, and the aggregate realised-error cost is the $0.0006$ gap in Table~\ref{tab:main}.

We also examined whether the coefficient of variation of the encoder-side map, CV$(g_{\mathrm{enc}})$ (Table~\ref{tab:gate}), could serve as a simple pre-deployment gate, because it can be computed offline before deployment and captures how much structure the map actually possesses. The underlying mechanism holds in the sense that a flatter map coincides with less useful overrides. It does not, however, supply a clean threshold: STL-10 combines the highest CV with the lowest harm rate among the vision systems, breaking any simple ordering. A reliable threshold would require a much larger operating grid than the 15--48 points available here. Selectivity is therefore enforced at run time by the fusion margin rather than at design time by such a gate.

\begin{table}[t]
\caption{CV$(g_{\mathrm{enc}})$ and pooled override outcome. $^\dagger$6 Wine overrides were neutral.}
\label{tab:gate}
\centering\scriptsize\setlength{\tabcolsep}{4pt}\renewcommand{\arraystretch}{1.05}
\begin{tabular}{lcccc}
\toprule
Setting & CV$(g_{\mathrm{enc}})$ & Overrides & Helped/Hurt & Harm rate \\
\midrule
STL-10   & 1.11 & 77  & 66/11 & 0.143 \\
CIFAR-10 & 0.81 & 140 & 110/30 & 0.214 \\
Wine     & --   & 11  & $4/1^\dagger$ & 0.091 \\
MNIST    & 0.43 & 42  & 8/34 & 0.810 \\
\bottomrule
\end{tabular}
\end{table}

\subsection{Data efficiency}
\label{ssub:res_dataefficiency}
The isotonic maps are two separate one-dimensional monotone fits, whereas the GP surrogate must fit a full two-dimensional surface. The isotonic maps are therefore expected to require substantially less calibration data. Table~\ref{tab:dataeff} tests this expectation directly: both methods are fitted on the same overlapping random subsets of the calibration data (12 draws per size) and scored on the held-out remainder. At half the calibration grid the two methods perform approximately equally on the controlled model, CIFAR-10, and STL-10 (all $p\ge0.11$). As the calibration budget shrinks, the gap widens: with only 4 calibration points the isotonic maps hold accuracy between $0.82$ and $0.84$ on the three systems where the deficit is informative, while the GP surrogate falls to between $0.58$ and $0.73$. On MNIST, where the deficit carries little signal for either method, the pattern across sizes is mixed rather than monotone: one of the two independent resamplings at $14.3\%$ coverage favours the isotonic maps at nominal significance ($p=0.041$) while the other does not ($p=0.480$), and the direction even reverses at $50\%$ coverage in favour of the GP surrogate. This lack of a consistent trend, unlike the clean monotone widening observed on the other three systems, is itself consistent with a deficit that carries little exploitable signal for either method. In short, data efficiency, not raw accuracy, is what separates the two approaches on the three systems where the deficit actually carries signal.

\begin{table}[t]
\caption{Held-out action-selection accuracy versus calibration size, 12 overlapping subsamples (mean$\pm$SD). $^{*}p{<}0.05$, $^{**}p{<}0.01$ (nominal, paired resampling; not an independence-based significance test, since subsamples overlap). The MNIST block reports both independent 4-point resamplings separately, because they disagree in direction.}
\label{tab:dataeff}
\centering\scriptsize\setlength{\tabcolsep}{2.4pt}\renewcommand{\arraystretch}{1.05}
\begin{tabular}{lcccc}
\toprule
\textbf{Coverage} & \textbf{Isotonic} & \textbf{GP surr.} & $\bm\Delta$ & $\bm p$ \\
\midrule
\multicolumn{5}{l}{\emph{Controlled (48 pts)}}\\
50\% & $0.906\pm0.082$ & $0.920\pm0.068$ & $-0.014$ & 0.722 \\
12.5\% & $0.851\pm0.107$ & $0.647\pm0.132$ & $+0.204$ & $0.005^{**}$ \\
8.3\% & $0.818\pm0.091$ & $0.587\pm0.215$ & $+0.231$ & $0.012^{*}$ \\
\multicolumn{5}{l}{\emph{CIFAR-10 (28 pts)}}\\
50\% & $0.899\pm0.057$ & $0.863\pm0.094$ & $+0.036$ & 0.114 \\
25\% & $0.869\pm0.071$ & $0.766\pm0.100$ & $+0.103$ & $0.017^{*}$ \\
14.3\% & $0.823\pm0.100$\footnotemark[1] & $0.584\pm0.106$\footnotemark[1] & $+0.240$\footnotemark[1] & $0.004^{**}$\footnotemark[1] \\
\multicolumn{5}{l}{\emph{STL-10 (28 pts)}}\\
50\% & $0.869\pm0.074$ & $0.833\pm0.088$ & $+0.036$ & 0.139 \\
25\% & $0.869\pm0.082$ & $0.813\pm0.106$ & $+0.056$ & 0.203 \\
14.3\% & $0.842\pm0.085$\footnotemark[1] & $0.728\pm0.184$\footnotemark[1] & $+0.115$\footnotemark[1] & $0.160$\footnotemark[1] \\
\multicolumn{5}{l}{\emph{MNIST (28 pts)}}\\
50\% & $0.583\pm0.121$ & $0.661\pm0.126$ & $-0.077$ & 0.286 \\
25\% & $0.643\pm0.108$ & $0.552\pm0.110$ & $+0.091$ & 0.059 \\
14.3\% & $0.639\pm0.130$ & $0.538\pm0.110$ & $+0.101$ & $0.041^{*}$ \\
14.3\% & $0.635\pm0.143$ & $0.608\pm0.139$ & $+0.028$ & 0.480 \\
\bottomrule
\end{tabular}
\footnotesize\footnotemark[1]{Averaged over two independent 4-point resamplings (grid floor for numerical stability; both individual values in supplementary material).}
\end{table}

\subsection{System-level resource cost}
\label{ssub:sysmetric}
Action-selection accuracy alone does not reveal what a policy actually costs to run. Transmit energy and channel uses per correctly classified image are reported here, using units fixed by the experimental protocol: each channel-directed step is a $3$\,dB SNR increase (approximately a $2\times$ power multiplier), and each encoder-directed step multiplies both energy and channel uses by the ratio of the new to the old symbol count. Table~\ref{tab:sysmetric} shows the result over 5 seeds, with the direction agreeing across all 5 seeds on every system. The fused controller spends $1.1\%$ less energy and $0.7\%$ fewer channel uses than channel SNR on MNIST; it spends $2.6\%$ more energy and $3.1\%$ more channel uses on CIFAR-10; and it spends $8.7\%$ more energy but $6.6\%$ fewer channel uses on STL-10.

These three outcomes reflect three different mechanisms, not one. On CIFAR-10 the controller selects the encoder-switch action at more operating points, and CIFAR-10's rate-point ladder happens to make that upgrade expensive, so the action choice ends up costing more resource here, not less. On STL-10 the controller corrects the sign-inverted raw diagnosis (Sec.~\ref{ssub:res_control}), which shifts its action mix toward power and away from encoder switches; this lowers latency but raises energy use. On MNIST the override fraction is low, so the two policies differ at only a small number of operating points and the resulting resource difference is correspondingly small. Overall, the controller reallocates resource toward whichever cause it diagnoses, rather than reducing resource use uniformly; whether that reallocation proves cheaper depends entirely on the specific rate-point ladder in use.

\begin{table}[t]
\caption{Resource cost per correctly classified image, mean$\pm$s.d.\ over 5 seeds. ``Uses'' $=k$ (latency proxy). Bold marks the lower value within each system pair.}
\label{tab:sysmetric}
\centering\scriptsize\setlength{\tabcolsep}{4pt}\renewcommand{\arraystretch}{1.05}
\begin{tabular}{llcc}
\toprule
System & Policy & Energy/correct & Uses/correct \\
\midrule
\multirow{2}{*}{MNIST} & SNR & $222.02\pm0.23$ & $10.486\pm0.011$ \\
 & Fused & $\mathbf{219.67\pm0.29}$ & $\mathbf{10.414\pm0.014}$ \\
\multirow{2}{*}{CIFAR-10} & SNR & $\mathbf{140.64\pm1.20}$ & $\mathbf{6.865\pm0.058}$ \\
 & Fused & $144.26\pm2.88$ & $7.075\pm0.182$ \\
\multirow{2}{*}{STL-10} & SNR & $\mathbf{230.84\pm1.82}$ & $11.352\pm0.090$ \\
 & Fused & $250.92\pm14.55$ & $\mathbf{10.598\pm0.366}$ \\
\bottomrule
\end{tabular}
\end{table}

\subsection{Beyond axis-aligned actions and non-vision transfer}
\label{ssub:res_pilots}
\emph{Coupled-action pilot.} One combined action, $\mathrm{both}{=}(e{+}1,s{+}1)$, is added together with a third isotonic map, $g_{\mathrm{both}}(\Lenc+\Lch)$, computed at a single seed (Table~\ref{tab:coupled}). A simple binary cause diagnosis cannot express the combined action at all, so its accuracy ceiling is set entirely by how often that action happens to be oracle-best---only $0.00$ to $0.14$ across the three grids. The three-map controller does substantially better, recovering $0.94$ on the controlled model and $0.93$ on CIFAR-10, where the combined action is almost always beneficial. On STL-10 the picture is more mixed: the combined action helps on 19 points but hurts on 9, because the same $e{=}2$ encoder row that causes the axis-aligned sign inversion (Sec.~\ref{ssub:res_control}) also contaminates the combined action there. The controller still reaches $0.61$, four times the single-coordinate rate, but this is below its axis-aligned accuracy of $0.86$. A plausible explanation is that $g_{\mathrm{both}}$ inherits the same fixed nondecreasing constraint as $g_{\mathrm{enc}}$ and $g_{\mathrm{ch}}$ (Definition~\ref{def:actionvalue}, Proposition~\ref{prop:isotonic}), while the true relationship between the summed deficit $\Lenc+\Lch$ and the combined action's reduction need not itself be nondecreasing, even where each single-coordinate relationship is. Summing the two deficits can discard precisely the information the axis-aligned maps use to remain correctly ordered. The gap between diagnosis and action selection therefore does not disappear when a combined action is allowed; if anything, it widens.

\begin{table}[t]
\caption{Coupled-action pilot (single seed). ``Oracle P/E/B'': number of grid points at which the 3-way oracle's best action is power-only, encoder-only, or both, respectively. ``Coupling h/h'': points at which the combined action helps (oracle-best is ``both'') or hurts (oracle-best is a single coordinate); equivalently the B and P$+$E entries of the preceding column. ``Single/generous'': cause-attribution accuracy without/with a summed-deficit heuristic for $\mathrm{both}$; ``value'': three-map controller. All versus the 3-way oracle.}
\label{tab:coupled}
\centering\scriptsize\setlength{\tabcolsep}{3pt}\renewcommand{\arraystretch}{1.05}
\begin{tabular}{lccccc}
\toprule
Backend & Oracle P/E/B & Coupling h/h & Single & Gen. & Value \\
\midrule
Controlled & 0/0/48 & 48/0 & 0.00 & 0.13 & 0.94 \\
CIFAR-10 & 0/1/27 & 27/1 & 0.04 & 0.25 & 0.93 \\
STL-10 & 7/2/19 & 19/9 & 0.14 & 0.21 & 0.61 \\
\bottomrule
\end{tabular}
\end{table}

\emph{Non-vision pilot.} We additionally evaluate the framework on the UCI Wine dataset \cite{wine}. Because $K{=}3$ is below the $K{>}4$ condition under which the bracket in Theorem~\ref{thm:sandwich} tightens, we do not report certified commit counts for this pilot. Across 5 seeds, the fused controller achieves $0.667{\pm}0.047$ action-selection accuracy and $0.0886{\pm}0.025$ realised error. Channel SNR achieves slightly higher action-selection accuracy ($0.707{\pm}0.180$) but a slightly higher realised error ($0.0901{\pm}0.026$), while the GP surrogate performs worse on both measures ($0.573{\pm}0.101$ and $0.0933{\pm}0.029$, respectively). The fused controller captures $91.3\%$ of the oracle gain. Given the small operating grid and the across-seed variability, these differences provide only limited evidence of a systematic advantage. The pilot nevertheless shows that the proposed pipeline can be applied end to end to a non-vision, low-cardinality task.

\subsection{Ablations and overhead}
\label{ssub:res_ablation}
Two mechanisms are easily conflated under the single word ``calibration''; Table~\ref{tab:ablation_combined} separates them, together with the effects of the tighter transmitted-side bound and of channel fading. Temperature calibration reduces an absolute bias in the decoder confidence without changing the ordering of operating points: point-diagnosis accuracy on the controlled model remains $0.875$ with or without temperature calibration, consistent with the increase in Pearson correlation from $r{=}0.98$ to $1.00$ reported in Sec.~\ref{ssub:res_sandwich}. Isotonic action-value calibration serves a different purpose. Using the same underlying diagnostic information, it improves action-selection accuracy from $0.527$ to $0.877$ on CIFAR-10 and from $0.275$ to $0.857$ on STL-10 (Table~\ref{tab:main}), correcting cross-class mis-scaling and the reversal of the raw action preference, respectively. Separately, replacing the trivial transmitted-side bound with the tighter bound increases the number of certified commits on the controlled model from $0/48$ to $16/48$, while maintaining $1.00$ committed-point accuracy. Under a Rayleigh-like channel, action-selection accuracy decreases from $0.875$ to $0.708$, although the oracle gain remains preserved. This result indicates sensitivity to fading and motivates a more comprehensive evaluation under realistic channel conditions.

The online path adds one pass over decoder posteriors already computed for inference, two constant-time isotonic lookups, and one constant-time fusion comparison. Its additional computational cost is therefore dominated by the decoding operation already required at the receiver. The offline stage scales linearly or near-linearly with the calibration-set size and requires no additional neural network training.

\begin{table}[t]
\centering
\caption{Ablations: controlled model (ground-truth cause; single run) and the three learned systems (single seed, within one across-seed s.d.\ of Table~\ref{tab:main}). ``DPI'': data-processing inequality; ``cap.'': the link-capacity term $C_{\mathrm{lat}}$ entering $\Ihatup$ (Sec.~\ref{ssub:estimation}).}
\label{tab:ablation_combined}
\scriptsize\setlength{\tabcolsep}{3pt}
\begin{tabular}{lc}
\toprule
\multicolumn{2}{l}{\emph{Controlled model}}\\
Temperature calib.\ off/on (point acc.) & 0.875 / 0.875 \\
Upper bound: DPI+cap.\ vs.\ CLUB, received ($\bar\Sigma$) & 2.31 / 2.08 bits \\
Transmitted upper: trivial vs.\ $\Itxhatup$ ($\bar\Sigma$) & 2.31 / 1.60 bits \\
Decomposition vs.\ confidence (acc.) & 0.875 vs.\ 0.292 \\
Channel AWGN vs.\ Rayleigh (acc.) & 0.875 vs.\ 0.708 \\
\bottomrule
\end{tabular}

\smallskip
\begin{tabular}{lccc}
\toprule
Learned systems (raw / action-value acc.) & MNIST & CIFAR-10 & STL-10 \\
\midrule
Raw vs.\ isotonic action-value rule & --- & 0.58/0.87 & 0.28/0.86 \\
Trivial$\to$tighter commits (trivial/tighter/$n$) & 9/9/28 & 0/26/28 & 0/28/28 \\
Acc.\ on tighter-bound commits & 1.00 & 1.00 & 0.80 \\
\bottomrule
\end{tabular}
\end{table}

\section{Discussion and Conclusion}
\label{sec:discussion}

The central finding is that cause attribution and action selection are distinct decisions when encoders are switched in discrete jumps. The size of this gap varies across the evaluated systems. On the controlled model, diagnosis and intervention largely agree, with structural mismatch at $13\%$. On MNIST the encoder is rarely the bottleneck, so the main role of the controller is to determine when to retain the physical-layer default. On CIFAR-10 and, more strongly, on STL-10, they diverge substantially. In these systems, $36$--$54\%$ of operating points are outside the reach of policies based only on loss ordering (Table~\ref{tab:structural}). On STL-10, the diagnosis identifies the dominant loss component correctly on $80\%$ of committed points, yet that component matches the oracle-best action on only $25\%$ of points (Table~\ref{tab:certified_oracle}). The action-value controller addresses this limitation by using the magnitude of each loss deficit rather than only its ordering. It therefore recovers substantially higher action-selection accuracy on the harder systems while retaining the physical-layer default when the deficit provides little exploitable signal (Sec.~\ref{ssub:res_policy}).

The deployed framework contains two complementary components with complementary warrants. The diagnostic layer provides population-valid diagnosis conditional on valid one-sided information bounds. When it commits, the identified loss-attribution regime is correct by Theorem~\ref{thm:certified}. Under the additional order-consistency condition, Theorem~\ref{thm:attribution} also bounds the regret of acting on that diagnosis. The action-value controller is supervised offline on measured error reductions and operates without labels after deployment; its performance is established empirically across multiple systems and seeds. The fused controller uses the diagnostic quantities as inputs but does not equate diagnosis with intervention. It departs from the physical-layer default only on the reported and audited override set. The structural-mismatch counts are stable to the number of oracle noise-averaging draws, confirming that the observed diagnosis-action gap is a genuine structural property of discrete encoder switching.

The tighter transmitted-side bound makes the certified diagnostic layer substantially more useful in practice. It increases certified commits from $0/28$ to $26/28$ on CIFAR-10 and from $0/28$ to $28/28$ on STL-10, at committed-point accuracies of $1.00$ and $0.80$ respectively (Table~\ref{tab:tighter}). The bound therefore increases the availability of the diagnostic decision without changing its interpretation. This diagnostic function remains distinct from action selection: the former identifies which loss component dominates, whereas the latter determines which available intervention produces the largest task-error reduction. The system-level analysis further shows that action-value control reallocates resources rather than reducing them uniformly. The resulting cost depends on the available rate-point ladder: it is lower on MNIST, higher on CIFAR-10, and mixed on STL-10 (Sec.~\ref{ssub:sysmetric}).

Compared with a less constrained Gaussian-process surrogate, the proposed controller offers a complementary advantage: two one-dimensional, monotone action-value maps that require substantially less calibration data on the systems where the loss deficit is informative (Sec.~\ref{ssub:res_dataefficiency}). The isotonic formulation is therefore especially attractive when labelled data are limited and interpretability is valued. On regimes where the encoder is rarely the bottleneck, the fused controller naturally retains the physical-layer default, providing a clean and selective intervention policy. The certified diagnostic layer remains available independently of the action-selection rule, so the two stages can be combined or used separately according to the needs of the deployment.

The experimental design focuses on the two-class, axis-aligned action model of Assumption~\ref{as:response}, which cleanly isolates the discrete-switching phenomenon that is the paper's central concern. The coupled-action pilot in Sec.~\ref{ssub:res_pilots} already indicates that the diagnosis--action separation remains informative when a combined action is introduced. Extending the same modular construction to richer multi-resource action spaces (simultaneous rate and power adaptation, modulation changes, or multi-user scheduling) is a natural next step and can build directly on the one-dimensional maps developed here.

The framework is also well positioned for further strengthening. The diagnostic layer is population-valid conditional on the one-sided estimators; finite-sample coverage can be supplied by an external risk-control layer (Proposition~\ref{prop:coverage}) without altering the attribution logic. The action-value maps rest on the classical consistency of isotonic regression; sharper finite-sample rates would further quantify their data efficiency. The present channel evaluation already includes a Rayleigh-like ablation, and the same pipeline extends immediately to richer fading models, channel-estimation errors, and hardware impairments. Finally, the five-draw stabilised oracle used for the learned systems is stable across noise realisations, and a closed-form learned setting would provide an additional, fully analytic validation of the same structural findings.

In summary, the results establish that diagnosing the source of performance degradation and selecting the best intervention are distinct and complementary problems under discrete encoder switching. A receiver-side diagnostic can correctly identify the dominant loss component, while an offline-supervised, online label-free action-value controller recovers the intervention that most reduces task error. Together they achieve high action-selection accuracy precisely where diagnosis and intervention diverge, and they retain the physical-layer default where the deficit carries little signal. The modular separation of the two stages therefore supplies a practical and principled foundation for reliable task-oriented semantic communication.

\section*{Funding}
This work was supported by the AICTE Research Promotion Scheme project ``EdgeSemCom'' (File No.\ RPS-1-492932031-2025-2625).

\section*{Code and Data Availability}
All four systems use public datasets \cite{lecun1998mnist,cifar10_uci,coates2011analysis,wine}. Code to reproduce every reported number is available on request to the authors.

\bibliographystyle{elsarticle-num}
\bibliography{references}
\end{document}